# A magneto-optic trap using a reversible, solid-state alkali-metal source

S. Kang[1,2,3,*], K. R. Moore[1], J. P. McGilligan[1,2], R. Mott[4], A. Mis[4], C. Roper[4], E. A. Donley[1] and J. Kitching[1]

[1]National Institute of Standards and Technology, Boulder Colorado, 80305, USA
[2]University of Colorado, Department of Physics, Boulder, Colorado, 80309, USA
[3]Key Laboratory of Atomic Frequency Standards, Wuhan Institute of Physics and Mathematics, Chinese Academy of Sciences, Wuhan, 430071, People's Republic of China
[4]HRL Laboratories, LLC; Malibu, California, 90265, USA
*kangsongbai@wipm.ac.cn

We demonstrate a novel way to form and deplete a vapor-cell magneto-optic trap (MOT) using a reversible, solid-state alkali-metal source (AMS) via an applied polarized voltage. Using ~100 mW of electrical power, a trapped-atom number of $5 \times 10^6$ has been achieved starting from near zero and the timescales of the MOT formation and depletion of ~1 s. This fast, reversible, and low-power alkali-atom source is desirable in both tabletop and portable cold-atom systems. The core technology of this device should translate readily to other alkali and alkaline-earth elements that could find a wide range of uses in cold-atom systems and instruments.

## I. Background

Laser cooling has revolutionized atom-based sensors and instrumentation. The low temperature of the atoms allows for long interaction periods and narrow spectroscopic linewidths that are critical for precision measurements. In all cold-atom systems, a source of (warm) atoms is required to provide an appropriate atom density for forming the MOT. Commercial rubidium (Rb) and cesium (Cs) alkali-metal dispensers (AMDs)[1] have been widely used in laboratory-based cold-atom experiments for decades due to their reliability and long lifetimes. However, these AMDs release alkali atoms via resistive heating, often requiring several watts during operation. Furthermore, dispensing via resistive heating is a non-reversible process; once the alkali metal is created, it cannot be recovered by the dispenser. This type of dispensing not only hinders the ability to precisely control the alkali-atom density for a cold-atom system over a large environmental temperature range, but also results in a long time constant for the alkali-atom density to decay, which negatively impacts the cycling rate, and hence performance, of cold-atom metrological experiments[2,3,4]. Such effects limit the use of AMDs in field-deployable, long-lifetime compact cold-atom sensors and clocks[5,6].

On the other hand, techniques like laser-ablated or current-pulsed AMDs[7,8,9] and light-induced atomic desorption (LIAD)[10,11,12,13] have been developed to considerably enhance the possibilities of modulating alkali-atom density in cold-atom systems. However, none of these techniques simultaneously meet the requirements of being fast, reversible, low-power and able to be miniaturized, which are important for developing a portable cold-atom physics package.

Recent efforts have focused on novel solid-state alkali sources that operate via an electrolysis process[14,15]. A promising candidate material for this process is a beta double-prime alumina ($\beta''$-alumina) ceramic, which features a high ionic mobility for alkali ions. By utilizing an applied voltage to control the flux of mobile alkali ions within the ceramic, this device has previously demonstrated bidirectional vapor-phase Rb sourcing and sinking functionality[16,17,18]. In this letter, we describe the novel implementation of a voltage-controlled, solid-state $\beta''$-alumina Rb AMS to demonstrate the formation and depletion of a Rb MOT in a vapor cell.

## II. Methods

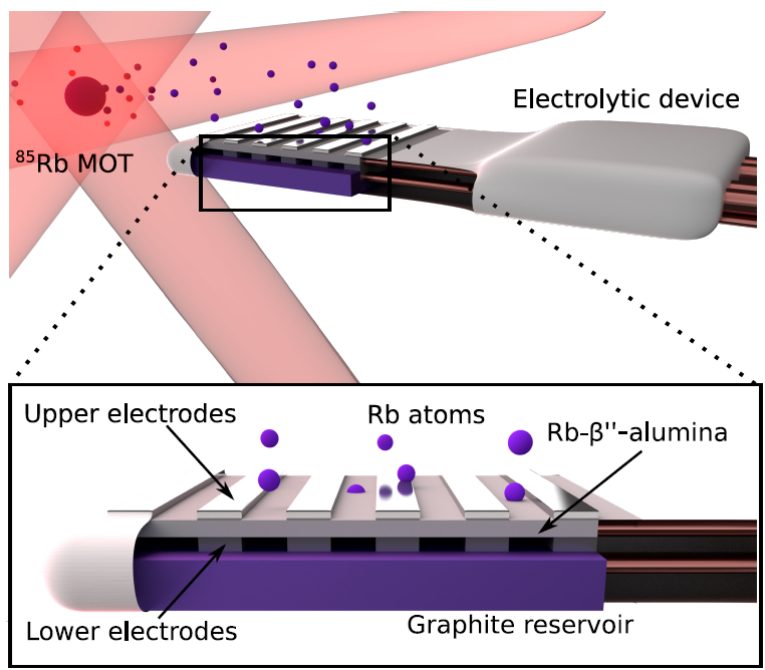

**FIG. 1**. Sketch of the experimental set-up with the six-beam MOT and in-vacuum electrolytic device. Inset: Illustrated cross-section of solid-state Rb AMS.

As shown in Fig.1, our experimental measurements are carried out in a standard six-beam MOT configuration[19,20]. The cooling and repumping light are derived from a single frequency-stabilized distributed-Bragg-reflector laser phase-modulated at 2.9 GHz. The carrier light is optically red-detuned by approximately two linewidths from the $^{85}$Rb $5S_{1/2}(F = 3) \rightarrow 5P_{3/2}(F' = 4)$ cooling transition and fiber-coupled to the experiment and collimated with a $1/e^2$ diameter of 3.8 mm and optical power of 4 mW in each of the six beams. Our ultra-high-vacuum system includes a standard 10-mm-by-10-mm cross-sectional-area vapor cell and a 2 L/s ion pump. The cell walls are coated with octadecyltrichlorosilane (OTS) to minimize alkali adsorption on the wall surface. Furthermore, a Rb AMD is installed in the vacuum system to initially load the AMS reservoir.

A CCD camera (timing resolution of 100 ms per data point) collects fluorescence images of the MOT with a spatial resolution of 17 $\mu$m. The trapped-atom number and the vapor-phase Rb density are estimated based on the MOT and background fluorescence intensity levels on the camera images, respectively.

The AMS is mounted inside the vapor cell about 1 cm from the intersection of the MOT beams. In the inset of Fig. 1, we show an illustrated cross-section of the AMS. The active surface area is 7.5 mm x 14.5 mm, and the thickness is approximately 2 mm. The solid-state electrolyte is sandwiched between grids of surface electrodes composed of fine Ti/Pt fingers with the top electrode having a period of 1 $\mu$m and with the lower electrodes in contact with a graphite reservoir. The high spatial density of the electrode fingers not only increases the atom-electrode contact area to aid electrochemical oxidation of Rb atoms, but also supports the efficient transport of ionizable atoms into the electrolyte. The lower electrodes and the graphite reservoir are sealed by a vacuum-compatible epoxy to isolate the stored Rb atoms from the cell environment. Heater wires buried inside the epoxy heat the AMS to approximately 100 °C with a power consumption of ~100 mW, which both increases the desorption rate of Rb atoms from the device surface and helps improve ionic mobility within the electrolyte. A voltage is applied through a 4.6 kΩ series resistor and the electrodes of the AMS. Depending on this voltage polarity, the AMS can reversibly source or sink Rb atoms from the vapor in the cell.

Prior to initial operation, the Rb AMD is operated for several hours to fill the vapor cell with Rb vapor. Then, a negative voltage is applied to the AMS for half an hour to load the graphite reservoir with Rb. After this Rb pre-loading process, we turn off the AMD and wait for the Rb density in the vapor cell to decay to a level such that the MOT signal drops below the detection threshold of the camera.

## III. Results

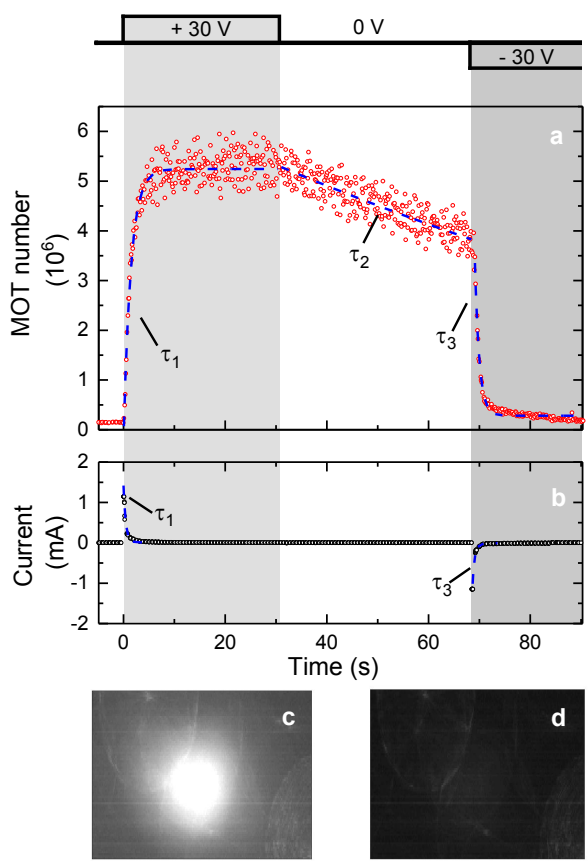

**FIG 2.** Demonstration of MOT formation and depletion using the device. **a**. The dynamic behavior of the trapped-atom number (red points) as the voltage on the device is changed. The number of trapped atoms is extracted from raw CCD image data by fitting to a 2D Gaussian function. **b.** The corresponding current flowing through the device (black points). Labelled gray and white regions indicate periods of different voltages applied to the AMS: 1) +30 V for $t_1 < t < t_2$; 2) 0 V for $t_2 < t < t_3$; and 3) -30 V for $t > t_3$. **c.** A CCD image of the MOT at ~ 20 s. **d.** A CCD image of the MOT at ~ 90 s.

In Fig. 2**a**, we demonstrate MOT formation and depletion controlled solely via the voltage applied across the Rb AMB. The demonstration is divided into three periods. First, the voltage is set to +30 V ($t_1 \leq t < t_2$). When the voltage is applied ($t = t_1$), the trapped-atom number increases from a near-zero baseline to a maximum level of $N_{MOT}$ ~5 x $10^6$ with a single time constant $\tau_1$ is ~1 s. An image of the MOT at this maximum fluorescence level is shown in Figure 2**c**. Next, the voltage is decreased to 0 V ($t_2 \leq t < t_3$). A slow decay of the number of trapped atoms is observed with a time constant $\tau_2$ ~100 s . Finally, the voltage is reversed to –30 V ($t \geq t_3$). The number of trapped atoms falls rapidly with a depletion time constant $\tau_3$ of ~1 s. An image of the MOT at the end of the cycle with the voltage reversed is shown in Figure 2**d**. The cooling beams and trap magnetic fields are on throughout. Because the non-Rb background pressure was ~1 × 10$^{-7}$ Torr, the time constants for the MOT loading process were less than 100 ms for all Rb vapor densities, and hence did not limit the dynamic behavior of the trapped atoms in our experiment.

We were able to monitor the background Rb density to some extent by measuring the change in the light captured on the CCD around the edges of the image where the cold atom fluorescence was low. We observed changes in the Rb density that roughly corresponded to the MOT number shown in Figure 2**a**, but with time constants $\tau_1$ and $\tau_3$ about 30% longer than the MOT time constants. This may suggest some weak saturation of the cold atom number at the higher background alkali densities.

The current, shown in Fig.2**b**, measures the transport of Rb ions across the electrolyte. Those Rb ions ultimately either plate out as Rb metal on the device surface or accumulate behind the electrodes. During the first period ($t_1 < t < t_2$), the current exhibits a rapid decay with a time constant of 0.4(1) s, likely due to electrode polarization, and then evolves into a slow decay. The positive initial spike (~1 mA at $t \sim t_1$) indicates a high Rb ion transport rate (~10$^{16}$ ions/s) from the graphite reservoir through the ionic conductor. If all of those Rb ions were to recombine with electrons and desorb as vapor-phase Rb atoms, that would yield a much higher Rb density (~10$^{15}$ /cm$^3$) in the cell volume (~10 cm$^3$) than is observed. Thus, it is likely that most of the Rb ions conducted through the electrolyte either accumulate behind the electrodes or form a Rb metal thin film on the AMS surface rather than immediately desorbing into the vapor. The subsequent low current (~10 $\mu$A for 2 s < t < $t_2$) likely indicates a slow rate of Rb-atom plating on the surface. The current during the third period ($t_3 < t$) exhibits a negative initial spike (~ -1 mA at $t \sim t_3$) with a decay time constant of 0.4(1) s followed by a low negative current (~ -10 $\mu$A at 70 s < t). The negative current spike indicates a high Rb-ion transport rate, presumably removing any Rb accumulated beneath the electrodes and most of the Rb metal film still on the surface. The low current is likely due to subsequent sinking from any remaining Rb metal film and from the vapor-phase Rb. The current dynamic ($\tau_1$ and $\tau_3$) is faster than the trapped-atom number and Rb vapor density, because of the additional Rb-atom desorption/adsorption process. The current time constants might be limited by the effective electrical contact resistance between the accumulated ions and the electrodes. This capability to rapidly sink Rb is a unique feature that distinguishes this electrochemical AMS from traditional AMDs. We note that the peak power for Rb sinking and sourcing is only ~10 mW, much lower than the heating power.

**V. Discussion and Summary**

This work presents a new mechanism for manipulating the vapor-cell cold-atom samples by utilizing a solid-state electrochemical Rb AMS. In Table I, we compare the characteristics of the vapor-cell MOT formed using different Rb sources. Using the AMS, a trapped-atom number of ~5x10$^6$ have been achieved from a near-zero MOT background. The maximum MOT number achievable could be further improved if the background gas pressure was lower. Such a high dynamic range can be obtained by the pulsed AMD technique but not by LIAD. Generally, there is a tradeoff between a high dynamic range and a fast depletion time for alkali-atom vapor pressure[21]. However, due to the unique active-reversible function of the AMS, the MOT depletion time constant can be as fast as ~1 s in our apparatus. V. Dugrain et. al. has obtained a 100 ms vapor-pressure modulation period in a pulsed AMD set-up, but that system requires an external, additional apparatus to thermally sink the alkali vapor[9]. Due to the electrochemical operating principle of the AMS, both sourcing and sinking functions are achieved by simply applying a voltage across the compact, in-vacuum package. Moreover, the AMS operates with a power of only ~100 mW limited by heating, which is critical for use in a portable, battery-powered instrument. The capacity of AMS could be as high as ~1 $\mu$g after a half an hour pre-loaded process which can guarantee a long lifetime for a cold-atom microsystem. Likewise, the planar design of the AMS is conducive to system integration and mass fabrication.

In summary, this compact device meets some requirements for field-deployable cold-atom systems potentially allowing MOT operation over a range of ambient temperatures. Meanwhile, the AMS could also find applications in experiments like those requiring evaporative cooling, where large numbers of atoms could be loaded into magnetic traps at high atomic density and, subsequently, long trap lifetimes could be obtained by lowering the density[21]. The micro-fabrication of this device complements mass production and implementation with other compact cold-atom devices to enable increased MOT-number control and stability in future quantum technologies[5,22,23]. The AMS pre-loaded with Rb is potentially able to replace the commercial AMDs in future, further simplifying vacuum assemblies and allowing for lower-power operation. Additionally, the core technology of the device should translate readily to other alkali and alkaline-earth elements that find a wide range of uses in cold-atom systems.

Table I. Characteristics of a vapor-cell MOT formed by different Rb sources.

|  | Dynamic range | MOT depletion time | Operation method | Power Consumption | Reversible operation |
|---|---|---|---|---|---|
| AMB | High | ~1 s | Voltage | ~100 mW | Yes |
| Pulsed AMD[8] | High | ~10 s | Current | ~1 W | No |
| LIAD[12] | Low | ~10 s | LED | ~10 W | No |

**Acknowledgement**

This material is based upon work supported by the Defense Advanced Research Projects Agency (DARPA) and Space and Naval Warfare Systems Center Pacific (SSC Pacific) (Contract No.

N66001-15-C-4027). The authors acknowledge DARPA program manager Robert Lutwak as well as Logan Sorenson, Matthew Rakher, Jason Graetz, John Vajo, Adam Gross, and Danny Kim of HRL Laboratories, LLC for useful discussions. We further acknowledge Florian Herrault, Geovanni Candia, Stephen Lam, Tracy Boden, Margie Cline, Ryan Freeman, and Lian-Xin Coco Huang for assistance with device fabrication. This work is a contribution of NIST, an agency of the U.S. government, and is not subject to copyright. J.P.M gratefully acknowledges support from the English Speaking Union and Lindemann Fellowship.

**Author contributions statement**

S.K., K.R.M. and J.P.M. performed the experiment and analyzed the data together with crucial inputs from J.K. and E.A.D. C.S.R. conceived and designed the alkali metal source. R.P.M., A.M., and C.S.R. developed and fabricated the alkali metal source. S.K., K.R.M. and J.P.M. prepared the manuscript with contribution from J.K., E.A.D. and C.S.R.

**Competing interests**

The authors declare no competing interests.